\documentclass{nature}
\usepackage{epsfig}
\usepackage{amsmath}
\usepackage{amsmath,mathrsfs,amsbsy,color,graphicx,bm,amsthm,amsfonts}
\usepackage{units}
\usepackage{bbm}
\usepackage{times}
\usepackage{dcolumn}
\usepackage{mathrsfs}
\usepackage{amsmath,amssymb,epsfig,float}

\begin{document}

\title{Unconventional pairings of spin-orbit coupled attractive degenerate
Fermi gas in a one-dimensional optical lattice}
\author{Junjun Liang$^{1}$, Xiaofan Zhou$^{1}$, Pak Hong Chui$^{2}$, Kuang
Zhang$^{1}$, Shi-jian Gu$^{2}$, Ming Gong$^{2,*}$, Gang Chen$^{1,*}$,
Suotang Jia$^{1}$}
\maketitle

\baselineskip=0.75 cm

\begin{affiliations}
\item
State Key Laboratory of Quantum Optics and Quantum Optics Devices, Institute of Laser spectroscopy, Shanxi University, Taiyuan 030006, P. R. China \\
\item
Department of Physics and Center for Quantum Coherence, The Chinese University of Hong Kong, Shatin, N.T., Hong Kong, China \\ \newline
$^*$Corresponding authors, e-mail: skylark.gong@gmail.com or chengang971@163.com\newline
\end{affiliations}

\begin{abstract}
Understanding novel pairings in attractive degenerate Fermi gases is crucial
for exploring rich superfluid physics. In this report, we reveal
unconventional pairings induced by spin-orbit coupling (SOC) in a
one-dimensional optical lattice, using a state-of-the-art density-matrix
renormalization group method. When both bands are partially occupied, we
find a strong competition between the interband
Fulde-Ferrell-Larkin-Ovchinnikov (FFLO) and intraband
Bardeen-Cooper-Schrieffer (BCS) pairings. In particular, for the weak and
moderate SOC strengths, these two pairings can coexist, giving rise to a new
phase called the FFLO-BCS phase, which exhibits a unique three-peak
structure in pairing momentum distribution. For the strong SOC strength, the
intraband BCS pairing always dominates in the whole parameter regime,
including the half filling. We figure out the whole phase diagrams as
functions of filling factor, SOC strength, and Zeeman field. Our results are
qualitatively different from recent mean-field predictions. Finally, we
address that our predictions could be observed in a weaker trapped potential.%
\newline
\end{abstract}

\baselineskip=0.8 cm

Ultracold atoms have become standard toolboxes for simulating fundamental
physics with strong interactions\cite{Jaksch,Bloch}. Recently, these systems
are used to mimic the spin-orbit coupling (SOC)\cite{VG13}, which is one of
the most intriguing interaction in nature. In particular, the
one-dimensional (1D) SOC---the simplest non-Abelian gauge potential\cite%
{JD11}---has been realized experimentally in fermionic $^{40}$K\cite%
{PW12,RAW13,ZF14} and $^{6}$Li\cite{LWC12} atoms, by using a similar scheme
achieved in bosonic $^{87}$Rb atom\cite{YJL11}. This remarkable progress
opens an immediate possibility for exploring nontrivial quantum phases of
degenerate Fermi gases\cite{ZH14}. Some intriguing phases, including the
topological Bardeen-Cooper-Schrieffer (BCS)\cite%
{MG11,JZ11,MG12,KS12,RW12,XIJL12,MI13,HH13} and topological
Fulde-Ferrell-Larkin-Ovchinnikov (FFLO)\cite{CQU13,WZ13,XJL13} phases, have
been revealed. The defects in these topological phases are expected to host
self-Hermitian Majorana fermions, which are the major building blocks for
achieving topological quantum computation\cite{CN08}. The basic picture for
realizing these nontrivial topological superfluids is that SOC, Zeeman
field, and $s$-wave interaction can induce triplet $p$-wave pairing, when
the chemical potential just occupies one single band\cite{LPG01,CWZ08,Sato09}%
. In this regard, understanding the true pairing(s) in the
spin-orbit-coupled systems is essential for achieving these novel phases.

All the previous predictions, both in free space and optical lattice, are
demonstrated in the framework of mean-field theory\cite%
{MG11,MG12,MI13,JZ11,KS12,RW12,XIJL12,HH13,CQU13,WZ13,XJL13}. In the
detailed calculations, the pairing is simply assumed to take place between
two fermions with a total center-of-mass momentum $Q$, which serves as a
parameter to minimize the total free energy. $Q=0$ and $Q\neq 0$ correspond
to the BCS and FFLO pairings, respectively. This fundamental picture is also
widely used even in 1D systems\cite{XJL13,CC13,QU1D}. In fact, in 1D the
effect of quantum fluctuation becomes significant and the mean-field results
are, in principle, unreliable\cite{XWG13}. This means that the true pairings
in this new platform need to be examined more seriously, which is, however,
still lacking. This work is devoted to addressing this fundamental issue in
a 1D spin-orbit coupled optical lattice, using a state-of-the-art density
matrix renormalization group (DMRG) method\cite{SRW92,SU05}.

Our numerical results demonstrate that the relevant physics in this model is
completely modified by the SOC-induced triplet pairing\cite%
{LPG01,CWZ08,Sato09}. (I) When both bands are partially occupied, the SOC
can lead to a strong competition between the interband FFLO and intraband
BCS pairings, due to the induced momentum-dependent spin polarizations. (II)
For the weak and moderate SOC strengths, these two pairings can coexist,
leading to a new phase called the FFLO-BCS phase. This new phase is
characterized by a unique three-peak structure in pairing momentum
distribution. (III) For the strong SOC strength, the system is dominated by
the intraband BCS pairing in the whole parameter regime, including the half
filling. (IV) We figure out the whole phase diagrams as functions of filling
factor, SOC strength, and Zeeman field, in terms of the properties of
pairing correlations in both real and momentum spaces. All the results
predicted are qualitatively different from the recent mean-field predictions%
\cite{QU1D}. (V) Finally, we address the effect of the trapped potential on
pairing correlations and local density. We show that our predictions could
be observed in a weaker trapped potential, which is easily prepared in
experiments. \newline

\section*{{\protect\LARGE \textbf{Results}}}

\subsection{Model and Hamiltonian.}

We consider the following 1D Fermi-Hubbard model with a synthetic SOC\cite%
{CC13,QU1D}:
\begin{eqnarray}
\mathcal{H} &=&-t\sum_{l,s=\uparrow ,\downarrow }(c_{ls}^{\dagger }c_{l+1s}+%
\text{H.c.})+h\sum_{l}(n_{l\uparrow }-n_{l\downarrow
})+U\sum_{l}(n_{l\uparrow }-{\frac{1}{2}})(n_{l\downarrow }-{\frac{1}{2}})
\notag \\
&&+\lambda \sum_{l}(c_{l\uparrow }^{\dagger }c_{l+1\downarrow
}-c_{l\downarrow }^{\dagger }c_{l+1\uparrow }+\text{H.c.}),  \label{H1}
\end{eqnarray}%
where $c_{ls}^{\dagger }$ and $c_{ls}$ are the creation and annihilation
operators, with spin $s=\uparrow ,\downarrow $ (encoded by the hyperfine
states), at lattice site $l$, $n_{ls}=c_{ls}^{\dagger }c_{ls}$ is the number
operator, $t$ is the spin-independent hopping, $h$ is the Zeeman field along
$z$ direction, $U$ is the on-site attractive interaction, $\lambda $ is the
SOC strength, and \text{H.c.} denotes the Hermitian conjugate.

Recently, the spin-orbit coupled Bose-Einstein condensate in a 1D optical
lattice has been realized experimentally\cite{CH14}. Using a similar
technique, the Hamiltonian (\ref{H1}) could also be achieved in 1D
degenerate Fermi gases\cite{YAL10,GP14}. Moreover, the corresponding
parameters can be tuned widely. For example, the 3D optical lattice can be
prepared by the interference of three pairs of counter-propagating laser
beams\cite{MG02}. The corresponding periodic potential is $V_{\text{lattice}%
}=V_{0}\cos ^{2}(k_{w}x)+V_{0}\cos ^{2}(k_{w}y)+V_{0}\cos ^{2}(k_{w}z)$,
where $V_{0}$ is the lattice depth, $k_{w}=\lambda _{w}/2\pi $ is the wave
vector, and $\lambda _{w}$ is wavelength. By further using a large harmonic
transverse confinement $V_{\text{2D}}=m\omega _{\bot }^{2}r^{2}/2$ in 3D
optical lattice, i.e., the 2D harmonic potential frequency $\omega _{\bot }$
is far larger than the trapped frequency $\omega _{z}$ along the
weakly-confining axis, the required 1D optical lattice can be generated\cite%
{YAL10,GP14}. In such case, the 1D effective interaction is described by\cite%
{MO98,HM05}
\begin{equation}
U(z)=-\frac{2\hbar ^{2}}{m_{0}a_{1\text{D}}}\delta (z),  \label{ON}
\end{equation}%
with the 1D $s$-wave scattering length
\begin{equation}
a_{1\text{D}}=-\frac{a_{\bot }^{2}}{2a_{3\text{D}}}(1-C\frac{a_{3\text{D}}}{%
a_{\bot }}),  \label{a1d}
\end{equation}%
where $C\simeq 1.46$, $a_{\bot }=(2\hbar /m_{0}\omega _{\bot })^{1/2}$, $a_{3%
\text{D}}$ is the 3D $s$-wave scattering length, and $m_{0}$ is the atomic
mass. Equations (\ref{ON}) and (\ref{a1d}) show that the 1D on-site
attractive interaction can be tuned by Feshbach resonance\cite{Chin}. In
addition, for $^{40}$K\cite{PW12} or $^{6}$Li\cite{LWC12} systems, two spin
states are chosen respectively as $\left\vert \uparrow \right\rangle
=\left\vert 9/2,-9/2\right\rangle $ and $\left\vert \downarrow \right\rangle
=\left\vert 9/2,-7/2\right\rangle $, or $\left\vert \uparrow \right\rangle
=\left\vert 3/2,-3/2\right\rangle $ and $\left\vert \downarrow \right\rangle
=\left\vert 3/2,-1/2\right\rangle $. By using a pair of counter-propagating
Raman lasers, the 1D SOC in the Hamiltonian (\ref{H1}) can also be realized%
\cite{PW12,RAW13,ZF14,LWC12,YJL11}. Moreover, the SOC strength can be tuned
through a fast and coherent modulation of the Raman beams\cite{ZY03}; see
also recent experiment\cite{IanS}. For a typical optical lattice, the strong
SOC strength, $\lambda \sim t$, can be achievable\cite{MGUP}.

Since the effect of quantum fluctuation in 1D becomes significant, here we
perform a state-of-the-art DMRG method to discuss the Hamiltonian (\ref{H1}%
). Notice that the similar Hamiltonian has been discussed by means of the
same method\cite{EMS11}. In their work, they focused on the effect of the
many-body interaction on topological phase and related Majorana fermions. In
their calculation, the pairing term is preassigned to be the BCS pairing,
i.e., no FFLO pairing can be driven from the many-body interaction. Here we
mainly explore fundamental pairings induced by SOC, including the FFLO and
BCS pairings.

\begin{figure}[t]
\centering\includegraphics[width = 6.3in]{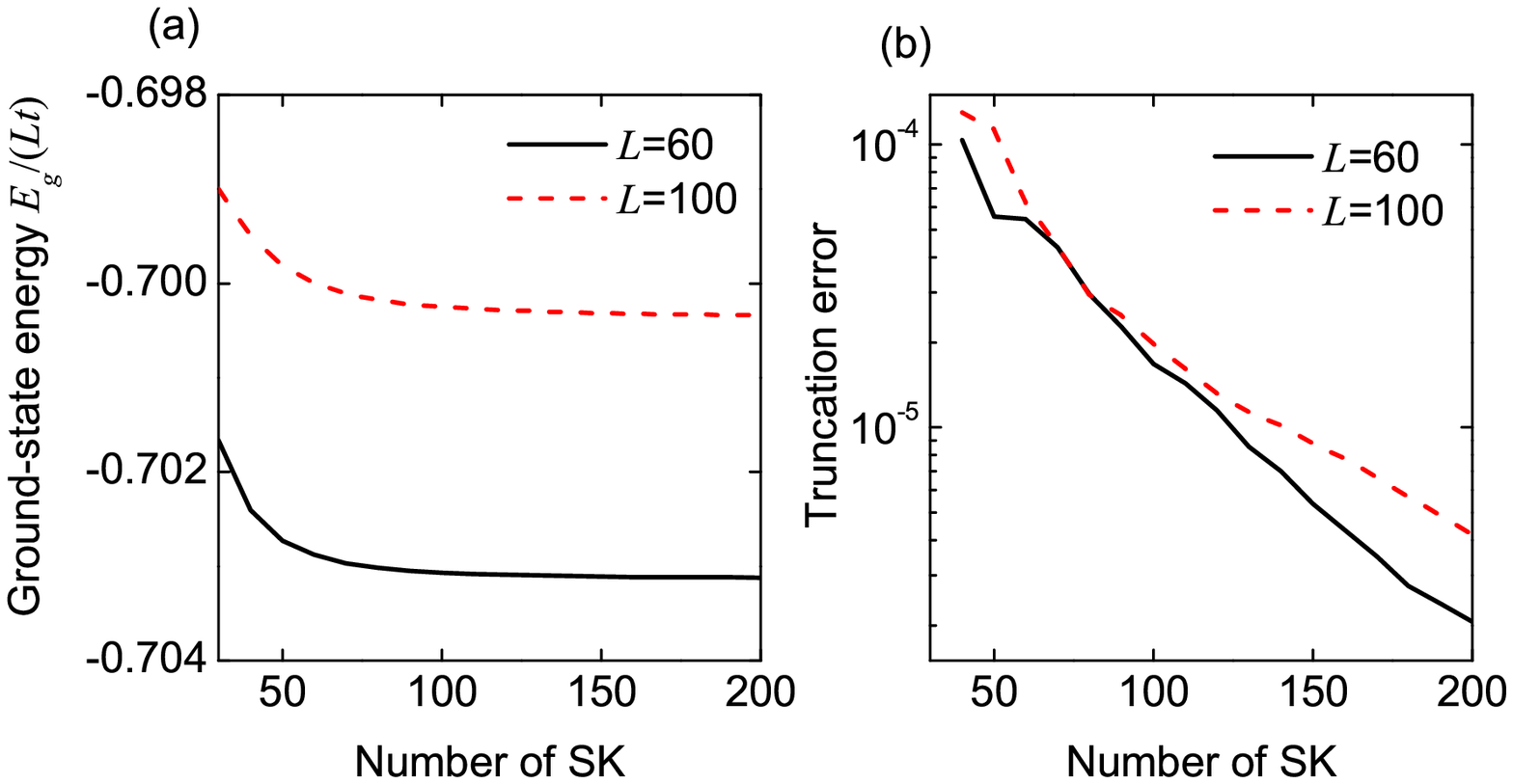}
\caption{\textbf{(a) The scaled ground-state energy }$E_{g}/(Lt)$\textbf{\
and (b) the truncation error as functions of the number of states kept (SK).}
}
\label{fig1}
\end{figure}

In the following calculations, the basic energy scale is chosen as $t=1$,
the on-site interaction is set to $U/t=-4$, and the lattice lengths are
chosen as $L=60$ and $100$. In addition, the open boundary condition is
taken into account and $20$ sweeps are always used. In Fig.~\ref{fig1}(a),
we plot the scaled ground-state energy $E_{g}/(Lt)$ as a function of the
number of states kept. It can be seen that the scaled ground-state energy $%
E_{g}/(Lt)$ tends to a stable value, when increasing the number of states
kept. This indicates that the number of states kept can be chosen as $150$
per DMRG block. In Fig.~\ref{fig1}(b), we show the truncation error as a
function of the number of states kept. When the number of states kept is
chosen as $150$, the truncation error is smaller than $10^{-5}$, which is
sufficient for the numerically-reliable results.

\subsection{Basic physical picture for unconventional pairings.}

\begin{figure}[t]
\centering\includegraphics[width = 3.0in]{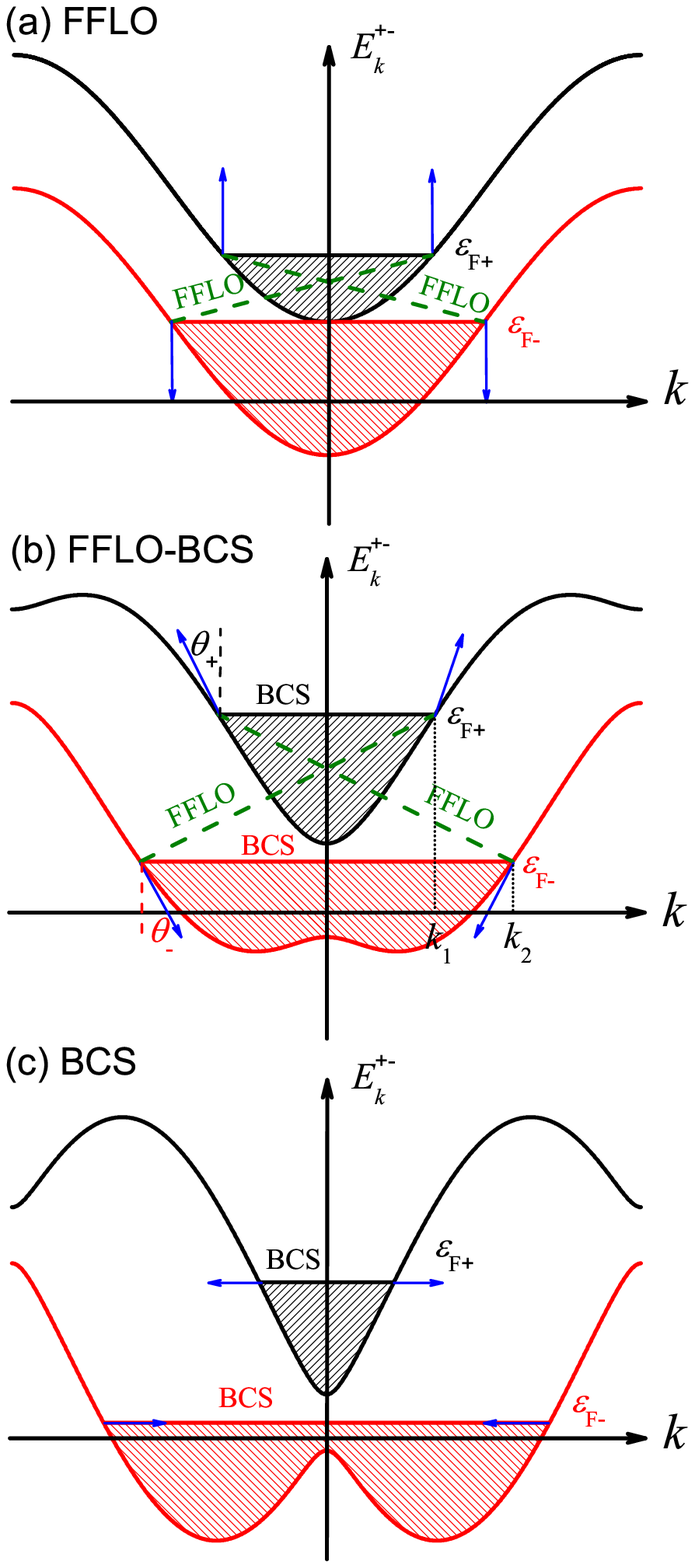}
\caption{\textbf{A schematic picture for illustrating unconventional
pairings in a 1D optical lattice.} }
\label{fig2}
\end{figure}

Before proceeding, we first illustrate the basic physical picture of the
Hamiltonian (\ref{H1}) with or without SOC. In the absence of SOC ($\lambda
/t=0$), the Hamiltonian $\mathcal{H}$ reduces to the well-studied
Fermi-Hubbard model\cite{ML07,TE10}, in which the Zeeman field breaks the
degeneracy of each band [the breaking of symmetry from $SO(4)$\cite{YangCN}
at the half filling and otherwise $SU(2)$ to $U(1)\otimes U(1)$]; however,
in both bands spin are still fully polarized along $z$ direction.
Consequently, the pairing can only be formed between two fermions at
different bands, due to the interspecies interaction. This pairing gives
rise to the well-known FFLO phase\cite{FF64,LO64}, which can be observed at
any nonzero population imbalance and has two nonzero center-of-mass momenta;
see Fig.~\ref{fig2}(a). However, this picture is completely modified by SOC,
since it leads to momentum-dependent spin polarizations
\begin{equation}
\mathbf{S}_{\pm }=\pm \frac{1}{\sqrt{4\lambda ^{2}\sin (k)^{2}+h^{2}}}%
(2\lambda \sin (k),0,h),  \label{Spin}
\end{equation}%
where $\pm $ denote the upper ($+$) and lower ($-$) bands. Since no spin
polarizations can be found in the $y$ component, the corresponding
spin-polarized angles can be defined, using only one variable, as
\begin{equation}
\tan (\theta _{\pm })=|\frac{2\lambda k_{\text{F}\pm }}{h}|.  \label{SPA}
\end{equation}%
When both bands are partially occupied, the physics is extremely
interesting. In this case, the SOC-induced triplet pairing can lead to the
intraband BCS pairing, which can compete with the original interband FFLO
pairing. This competition depends crucially on the spin-polarized angles $%
\theta _{\pm }$ at the Fermi points [in the mean-field level, the effective
pairing is $p$-wave type with pairing strength $\Delta _{\text{eff}%
}(k)\simeq \Delta \sin (\theta _{\pm })k$\cite{Chan}], as shown below.

For a weaker SOC strength, i.e., $|2\lambda k_{\text{F}\pm }|\ll |h|$, the
spin is still fully polarized along $z$ direction, and hence the
spin-polarized angles $\theta _{\pm }\sim 0$. It indicates that the
interband FFLO pairing dominates. When increasing the SOC strength, the spin
will gradually polarize towards the $x$ direction. For the weak and moderate
SOC strengths, the spin-polarized angles $\theta _{\pm }$ are typically of
the order of $\pi /10$ (see Supplementary Information). In such case, both
the intraband BCS and interband FFLO pairings are allowed and govern
simultaneously the true pairings of the Hamiltonian (\ref{H1}); see Fig.~\ref%
{fig2}(b). For the strong SOC strength, the spin is almost polarized along $%
x $ direction ($\theta _{\pm }\sim \pi /2$), and the intraband BCS pairing
thus dominates; see Fig.~\ref{fig2}(c). Therefore, we can expect a crossover
from the interband FFLO pairing to the intraband BCS pairing, when the
spin-polarized angles $\theta _{\pm }$ exceeds the critical values. We find
that this transition is nontrivial because these two pairings can coexist in
some parameter regimes; see below. Hereafter, all these pairings, which
arise from strong competition between the two pairing channels induced by
SOC and the Zeeman field, are called unconventional pairings. These
unconventional pairings can lead to rich superfluid phases, which can be
captured by considering pairing correlations in both real and momentum
spaces.

\subsection{Pairing correlation in real space.}

The pairing correlation function in real space is defined as\cite%
{AF07,AR08,MT08,MT10}
\begin{equation}
P(l,j)=\langle c_{l\downarrow }^{\dagger }c_{l\uparrow }^{\dagger
}c_{j\uparrow }c_{j\downarrow }\rangle .  \label{PC}
\end{equation}%
Without SOC, this pairing function can be used to identify the BCS and FFLO
pairings. Physically, for the FFLO pairing, $P(l,j)\sim \exp \left[ iQ(l-j)%
\right] $, which oscillates in real space and exhibits two peaks at $k=\pm Q$
in momentum space\cite{AF07,AR08,MT08,MT10}; see also discussions below. For
the BCS pairing, $Q=0$, and no oscillation can thus be found in real space.
In the presence of SOC, we also use this function as an important tool (but
not a unique tool) to identify the unconventional pairings of the
Hamiltonian (\ref{H1}).

\begin{figure}[t]
\centering\includegraphics[width = 6.3in]{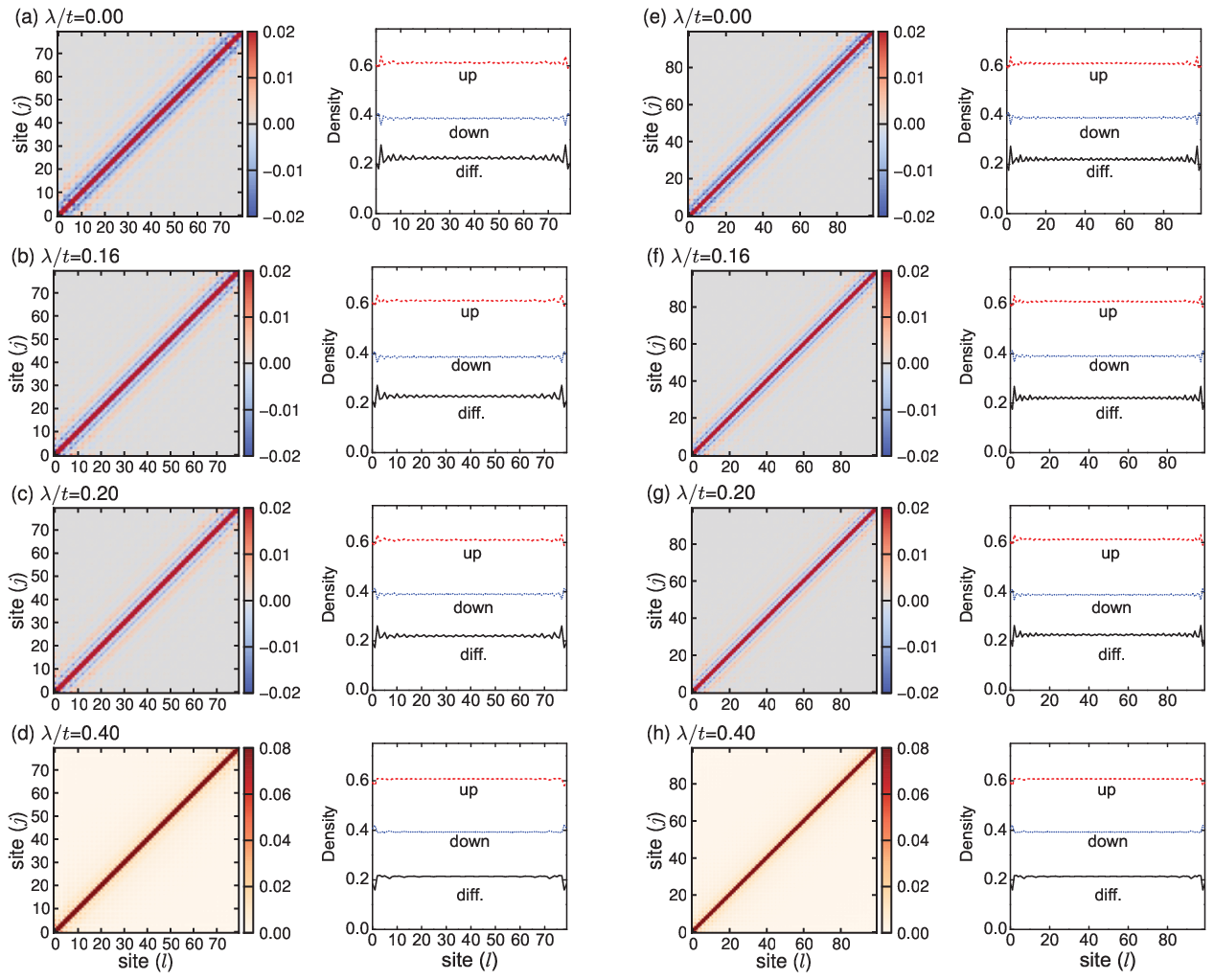}
\caption{\textbf{The pairing correlation functions }$P(l,j)$\textbf{\ and
local densities }$n(l)$\textbf{\ for the different SOC strengths.}}
\label{fig3}
\end{figure}

\begin{figure}[t]
\centering\includegraphics[width = 6.0in]{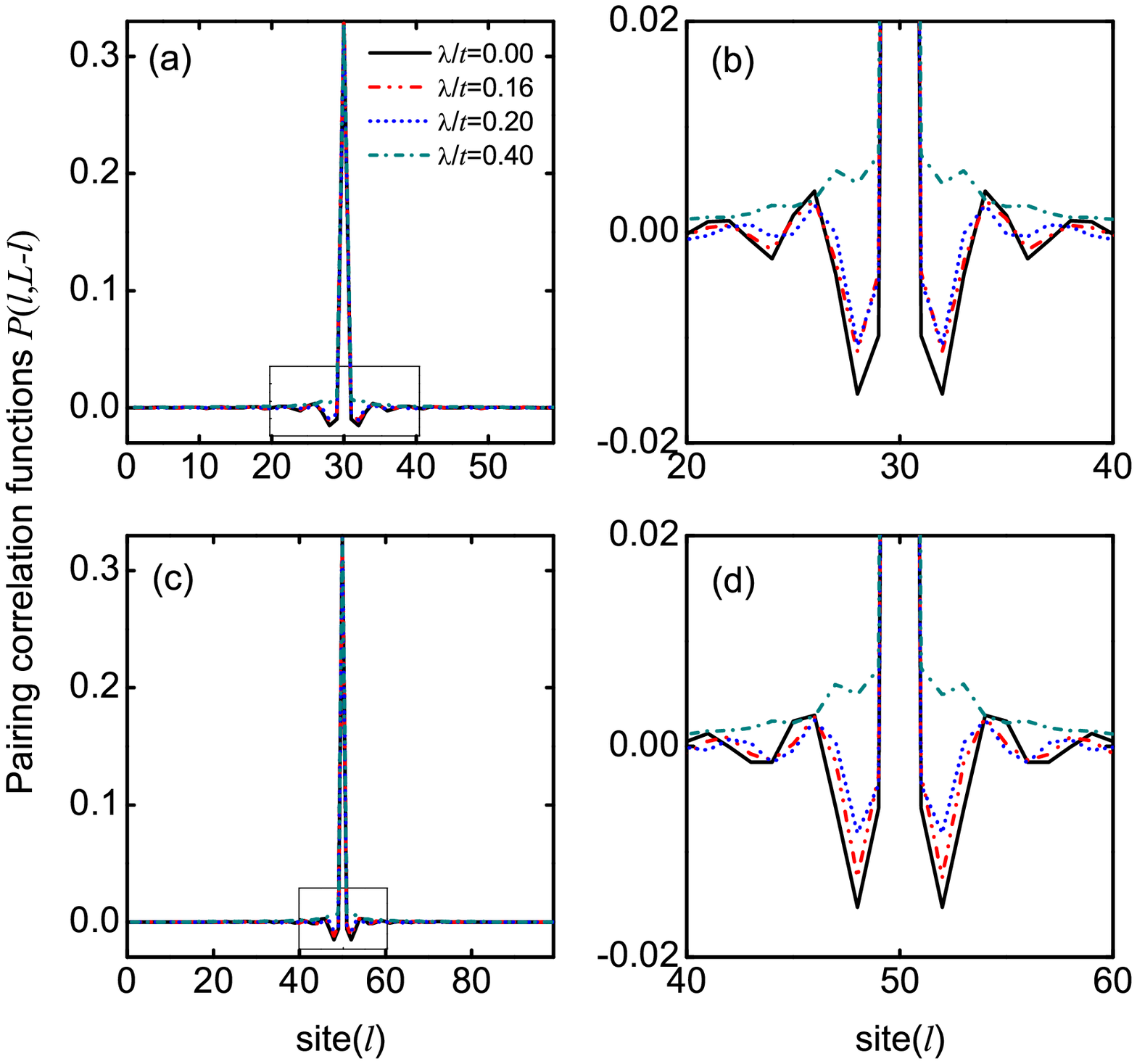}
\caption{\textbf{The off-diagonal pairing correlation functions }$P(l,L-l)$%
\textbf{.}}
\label{fig4}
\end{figure}

In Fig.~\ref{fig3}, we plot the pairing correlation functions $P(l,j)$ and
the local densities $n(l)$ for the different SOC strengths. Without SOC ($%
\lambda /t=0$), the pairing correlation in real space exhibits strong
oscillations in both magnitude and sign; see Fig.~\ref{fig3}(a). Moreover,
the local spin polarization also exhibits a similar oscillating behavior.
This indicates the emergence of a FFLO phase\cite{AF07,AR08,MT08,MT10}. For
the moderate SOC strength [see, for example, $\lambda /t=0.16$ and $\lambda
/t=0.2$ in Figs.~\ref{fig3}(b) and~\ref{fig3}(c)], the intraband BCS pairing
increases, and has a strong competition with the interband FFLO pairing. In
this case, the pairing correlations also have similar oscillating behaviors.
For the strong SOC strength [see, for example, $\lambda /t=0.4$ in Fig.~\ref%
{fig3}(d)], the pairing correlation exhibits a power decay with respect to $%
|l-j|$ without node, and no obvious oscillation of spin polarization can be
identified (the oscillation of spin polarization near the two ends is
attributed to the finite-size effect). This means that a BCS phase emerges,
as expected. In Fig.~\ref{fig4}, we plot the off-diagonal pairing
correlation functions $P(l,L-l)$ for the different SOC strengths. This
figure also shows clearly the oscillations of the pairing correlation in
real space, when the SOC strength is not very strong. This oscillation is
gradually suppressed by increasing the SOC strength.

\subsection{Pairing momentum distribution.}

Although the pairing correlation functions at the weak and moderate SOC
strengths exhibit the similar behaviors as those in the FFLO phase, their
corresponding pairing momentum distributions $P(k)$ have quite different
behaviors. The pairing momentum distribution---the Fourier transformation of
$P(l,j)$---is given by
\begin{equation}
P(k)=\frac{1}{2L}\sum_{l,j}P(l,j)e^{ik(l-j)}.  \label{PK}
\end{equation}

\begin{figure}[t]
\centering\includegraphics[width = 6.0in]{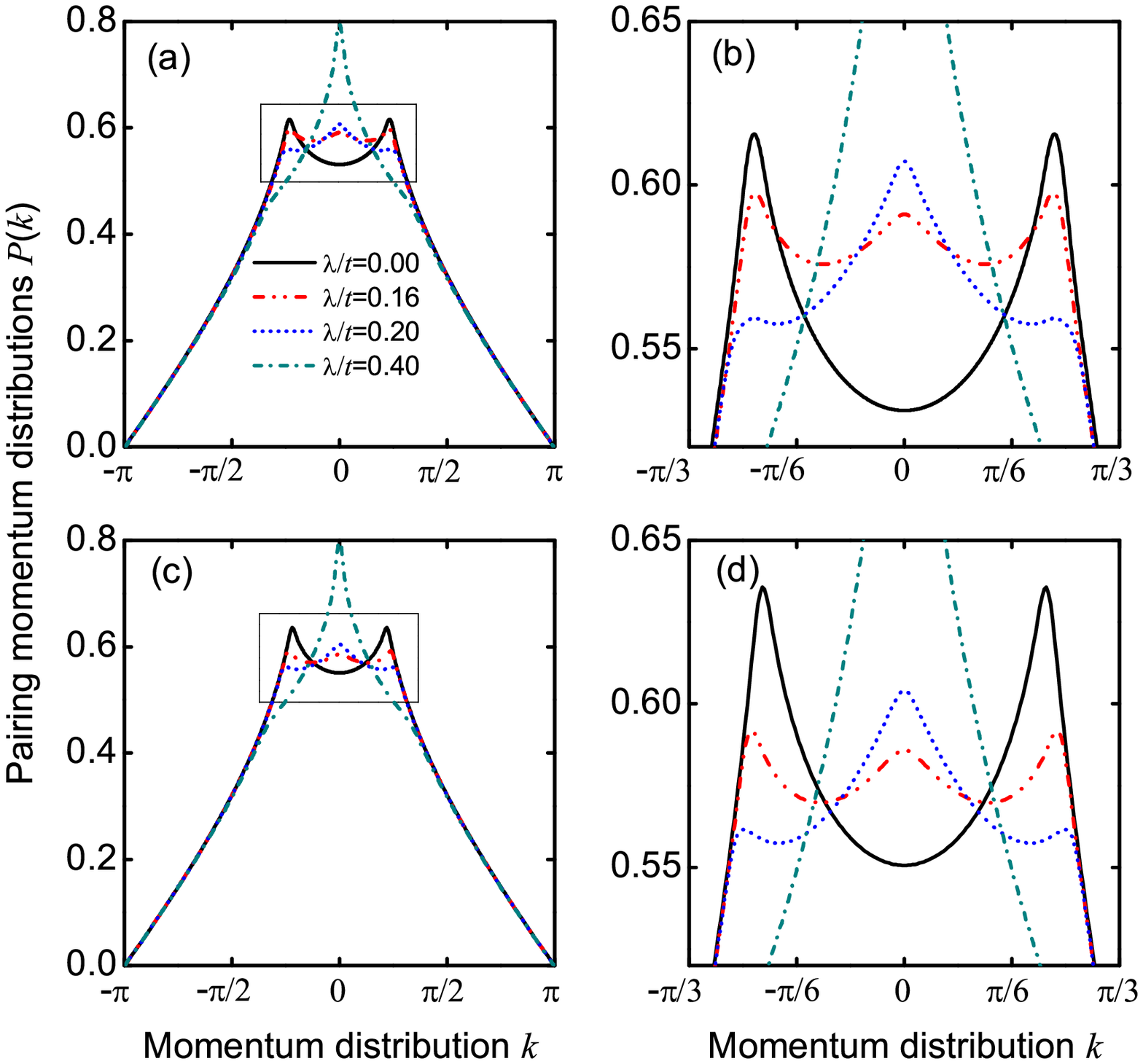}
\caption{\textbf{The pairing momentum distributions }$P(k)$\textbf{.}}
\label{fig5}
\end{figure}

Without SOC ($\lambda /t=0$), the polarization angles $\theta _{\pm }=0$,
and two nonzero center-of-mass momenta $\pm Q$($\neq 0$) can be found
explicitly; see Figs.~\ref{fig5}(a)-\ref{fig5}(b). This is a direct
consequence of inversion symmetry in our model, $P(k)=P(-k)$ is thus
expected. The corresponding phase is referred as the FFLO phase\cite%
{AF07,AR08,MT08,MT10}. When the SOC strength $\lambda /t=0.16$, the
polarization angles $\theta _{\pm }\simeq \pi /16$ (see Supplementary
Information). In such case, the dip at zero momentum of the pairing momentum
distributions $P(k)$ turns to a peak, while the other two peaks at $\pm Q$
change slightly (the detailed discussions are shown below). This indicates
that the pairing momentum distribution $P(k)$ has a unique three-peak
structure, which demonstrates clearly that the intraband BCS and interband
FFLO pairings can coexist. This result goes beyond the recent mean-field
prediction\cite{QU1D}. We call the corresponding phase the \emph{FFLO-BCS
phase}. When the SOC strength $\lambda /t=0.20$, the polarization angles $%
\theta _{\pm }\simeq \pi /13$, and the pairing mechanism is still similar to
that of $\lambda /t=0.16$. However, the peak of zero momentum is higher than
that of nonzero momenta, which implies that the intraband BCS pairing is
stronger than the interband FFLO pairing. For the strong SOC strength (see,
for example, $\lambda /t=0.4$), the polarization angles $\theta _{\pm }>\pi
/10$, and the intraband BCS pairing dominates. The corresponding phase is
referred as the BCS phase, in which the pairing momentum distribution $P(k)$
only has a peak at zero momentum. We need to emphasize that SOC affects
significantly the pairing momentum distribution $P(k)$ at the small momentum
regime. For the large momentum regime, the system's properties are
determined mainly by the short-range interaction, and the pairing momentum
distribution is thus unaffected by SOC; see also Figs.~\ref{fig5}(a) and \ref%
{fig5}(c).

Since the pairing momentum distribution $P(k)$ can be measured by the
time-of-flight imaging\cite{CAR04,FZ12}, the predicted three phases can be
observed directly in experiments. The corresponding boundary between the
FFLO and FFLO-BCS phases can be determined by
\begin{equation}
{\frac{d^{2}P(k)}{dk^{2}}}|_{k=0}=0,  \label{B1}
\end{equation}%
whereas the boundary between the FFLO-BCS and BCS phases can be determined
by
\begin{equation}
{\frac{dP(k)}{dk}}|_{k=Q}=0.  \label{B2}
\end{equation}

\begin{figure}[t]
\centering\includegraphics[width = 6.0in]{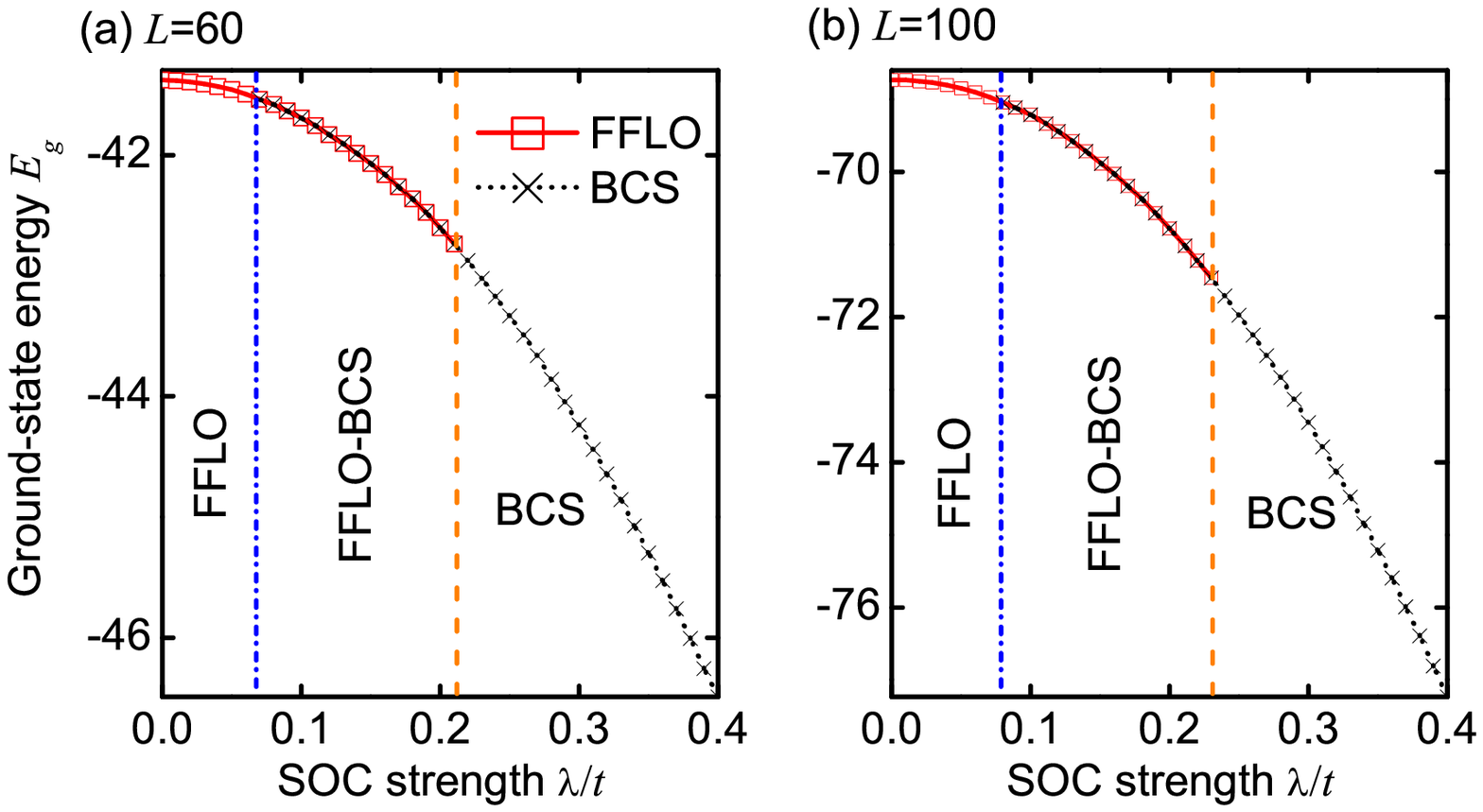}
\caption{\textbf{The ground-state energy }$E_{g}/t$\textbf{\ as a function
of the SOC strength.}}
\label{fig6}
\end{figure}

We now explain why the intraband BCS and interband FFLO pairings can coexist
in the FFLO-BCS phase. In the presence of SOC, there are two bands (see Fig.~%
\ref{fig2}), which contain spin-up and spin-down fermions. In the lower
band, there are lots of fermions, which, however, become less in the upper
band. Due to the existence of different spin-component fermions in the same
band, the pairings can, in principle, be formed in the same or different
bands, i.e., both the intraband BCS and interband FFLO pairings are allowed.
For the small spin-polarized angles, the interband FFLO pairing is favored,
while for the relative large spin-polarized angles, the intraband BCS
pairing is favored. More importantly, the corresponding ground-state
energies for both the intraband BCS and interband FFLO pairings are
degenerate in the FFLO-BCS phase (see Fig.~\ref{fig6}), which confirms the
coexistence of these pairings.

\begin{figure}[t]
\centering\includegraphics[width = 6.0in]{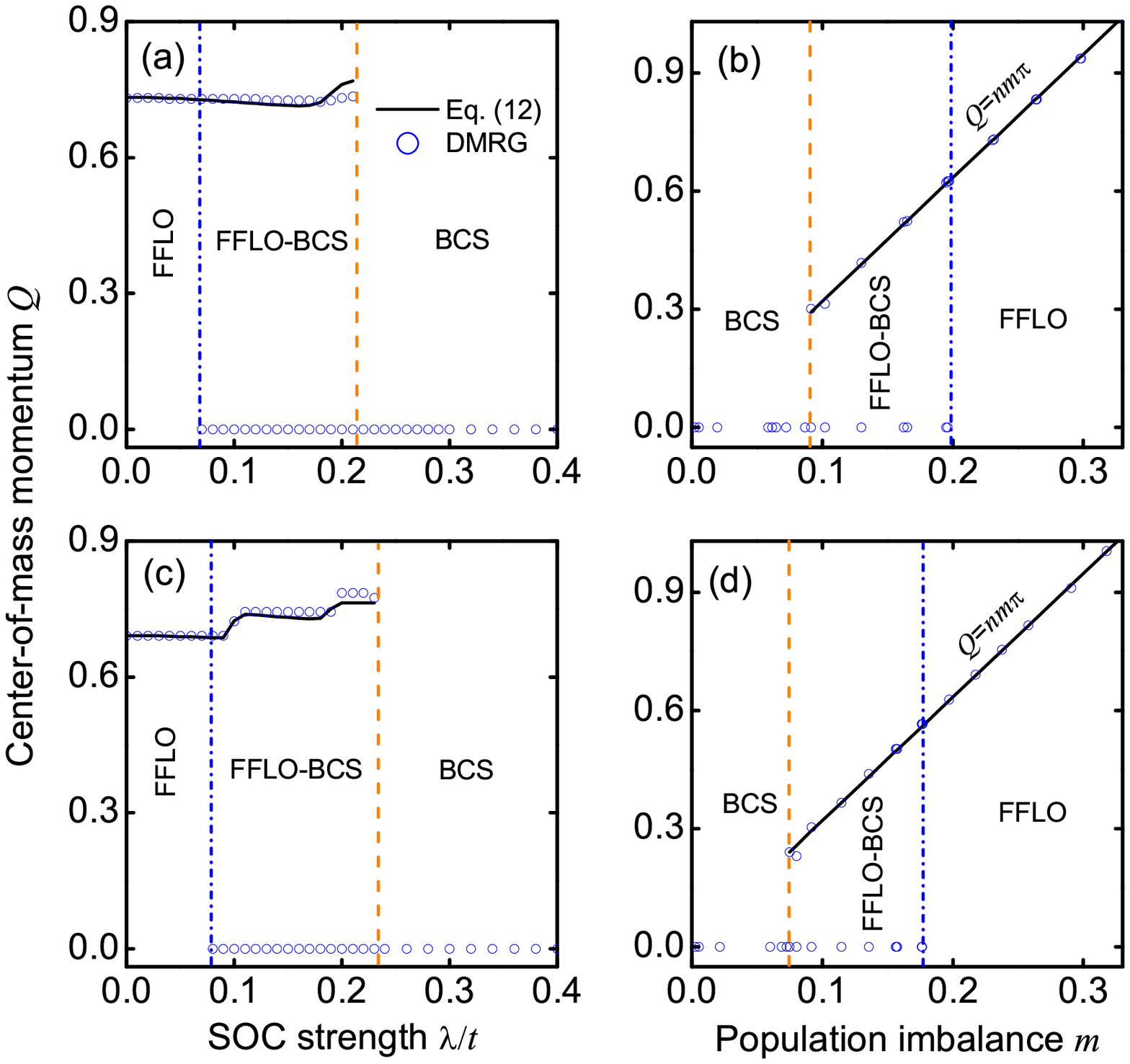}
\caption{\textbf{The center-of-mass momentum }$Q$\textbf{.}}
\label{fig7}
\end{figure}

In Figs.~\ref{fig7}(a) and \ref{fig7}(c), we plot the center-of-mass
momentum $Q$ as a function of the SOC strength $\lambda /t$. Numerically, $Q$
is determined by $dP(k)/dk=0$ and $d^{2}P(k)/dk^{2}<0$. We find that $Q$ is
a non-monotonic function of the SOC strength $\lambda /t$. Here we develop a
simple model to understand the relevant behavior. We assume that the Fermi
points for two bands have momenta $\pm k_{1}$ and $\pm k_{2}$, respectively.
These values are governed by the following equations:
\begin{equation}
n=\frac{(k_{1}+k_{2})}{\pi },\quad nm={\frac{1}{\pi }}\int_{k_{1}}^{k_{2}}%
\frac{h}{\sqrt{4k^{2}\lambda ^{2}+h^{2}}}dk,  \label{TEQ}
\end{equation}%
where $m=(N_{\uparrow }-N_{\downarrow })/N$ is the experimentally-measurable
population imbalance\cite{Martin2006,Guthrie2006}. The center-of-mass
momentum is determined by $Q=|k_{1}-k_{2}|$.

For simplicity, we adopt the simplified model in free space, with which the
analytical expression can be obtained perturbatively. We do not observe
quantitatively modification of our conclusion by replacing $k$ with $\sin
(k) $ for a lattice model. For the weak SOC strength, we employ the Taylor
expansion of Eq.~(\ref{TEQ}) (up to the leading term) to obtain
\begin{equation}
nm\pi =Q\left[ 1+{\frac{2(k_{1}^{3}-k_{2}^{3})}{3Qh^{2}}}\lambda ^{2}+{\frac{%
6(k_{2}^{5}-k_{1}^{5})}{5h^{4}Q}}\lambda ^{4}\right] ,  \label{CEM}
\end{equation}%
where $k_{1}=(n\pi -Q)/2$ and $k_{2}=(n\pi +Q)/2$. We assume the solution of
$Q$ has the following term
\begin{equation}
Q=nm\pi \left[ 1+\mathcal{A}_{2}\lambda ^{2}-\mathcal{A}_{4}\lambda ^{4}+%
\mathcal{O}(\lambda ^{6})\right] .  \label{eq-Q}
\end{equation}%
If letting the coefficient of $\lambda ^{2}$ and $\lambda ^{4}$ to be zero
by the Taylor expansion of Eq.~(\ref{CEM}), we can immediately find
\begin{equation}
\mathcal{A}_{2}=\frac{(3+m^{2})n^{2}\pi ^{2}}{6h^{2}}>0,  \label{AT}
\end{equation}%
\begin{equation}
\mathcal{A}_{4}=\frac{(15+50m^{2}-m^{4})n^{4}\pi ^{4}}{120h^{4}}>0.
\label{AF}
\end{equation}%
We find that Eqs.~(\ref{eq-Q})--(\ref{AF}) can well describe the evolution
of the center-of-mass momentum $Q$ in the presence of a weak SOC; see Figs.~%
\ref{fig7}(a) and \ref{fig7}(c). Moreover, without SOC ($\lambda /t=0$),
Eq.~(\ref{eq-Q}) reduces to the well-known result\cite{AF07,AR08,MT08,MT10}:
$Q=nm\pi $.

\begin{figure}[t]
\centering\includegraphics[width = 6.0in]{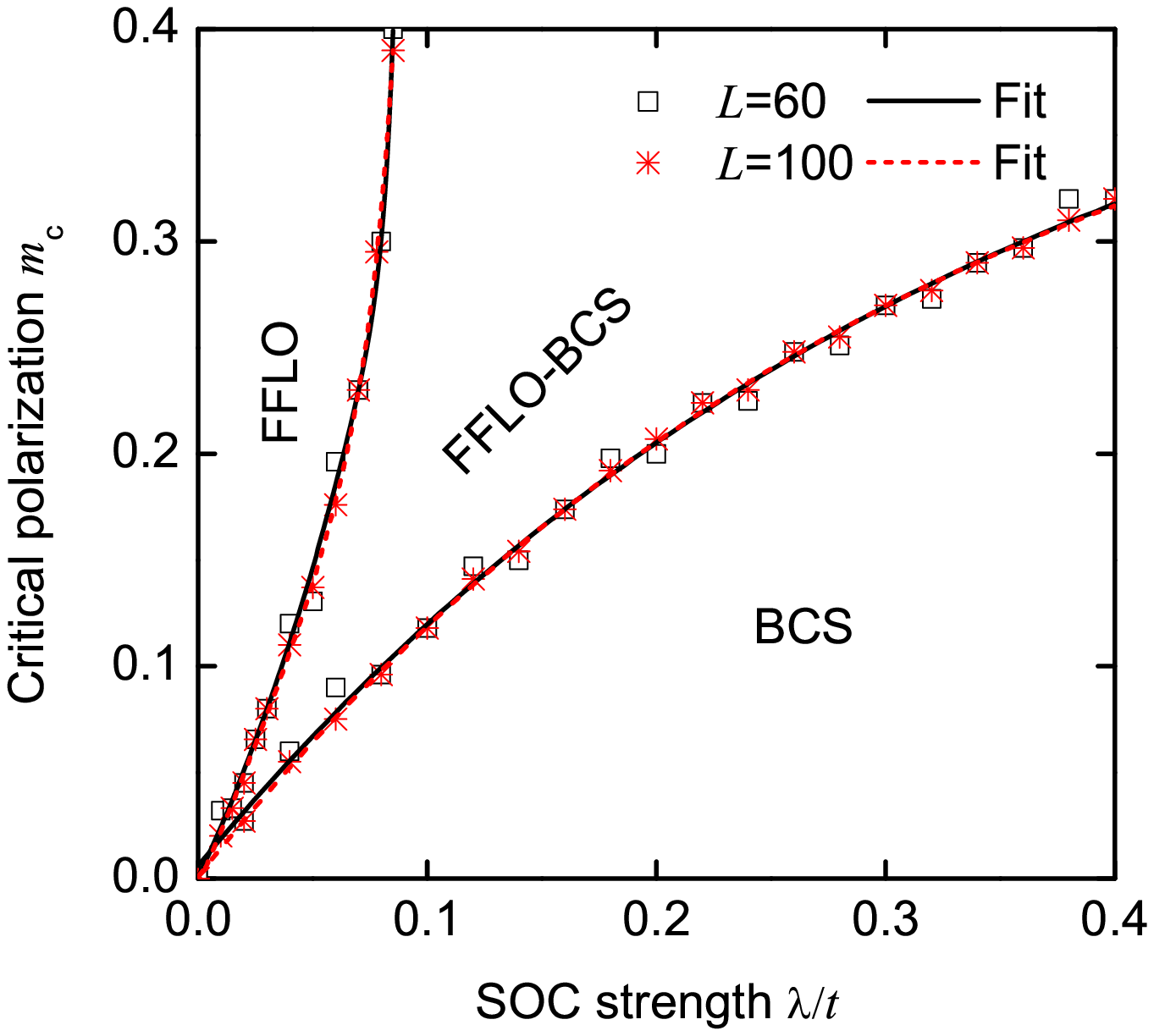}
\caption{\textbf{The critical population imbalance }$m_{c}$\textbf{\ as a
function of the SOC strengths}.}
\label{fig8}
\end{figure}

From Eq.~(\ref{eq-Q}), we also see that without SOC, any nonzero population
imbalance can give rise to the FFLO phase\cite{AF07,AR08,MT08,MT10}.
However, this basic conclusion is completely modified by SOC. In Figs.~\ref%
{fig7}(b) and \ref{fig7}(d), we plot the center-of-mass momentum $Q$\ as a
function of the population imbalance $m$, when the SOC strength $\lambda
/t=0.06$. We find that a finite population imbalance is required to realize
the interband FFLO pairing in our model. Moreover, in both the FFLO-BCS and
FFLO phases, $Q=nm(\lambda )\pi $, where $m(\lambda )$ is obtained from the
state-of-the-art DMRG calculations. In Fig.~\ref{fig8}, we plot the
relationship between the critical population imbalance $m_{c}$ for the
different phases and the SOC strength. Obviously, $m_{c}(\lambda =0)=0$, as
expected.

\begin{figure}[t]
\centering\includegraphics[width = 6.0in]{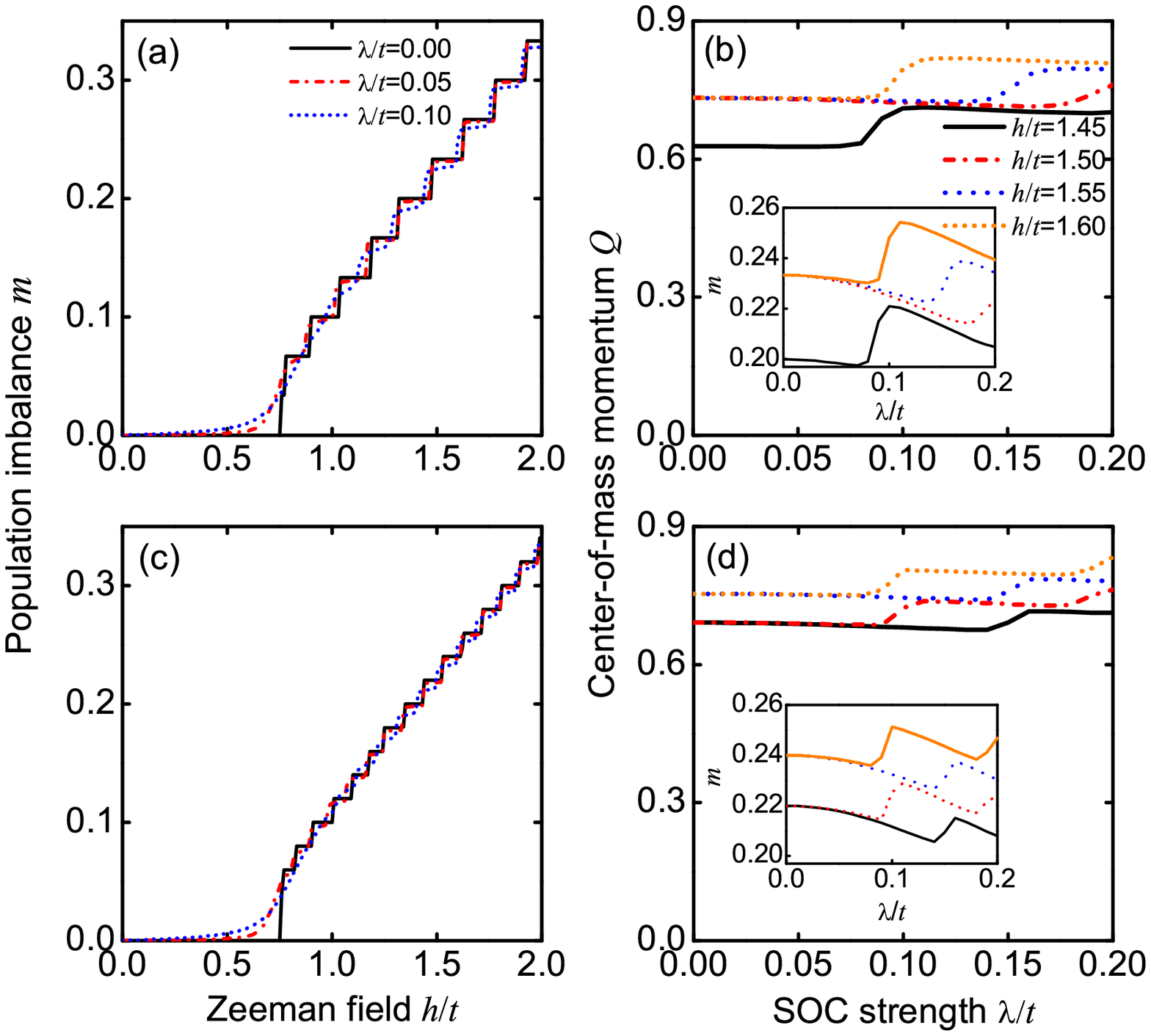}
\caption{\textbf{The population imbalance }$m$ \textbf{and the
center-of-mass momentums }$Q$ \textbf{for the different Zeeman fields.}}
\label{fig9}
\end{figure}

In Figs.~\ref{fig9}(a) and~\ref{fig9}(c), we plot the population imbalances $%
m$ as functions of the Zeeman field for the different SOC strengths, when $%
L=60$ and $L=100$. In the absence of SOC ($\lambda /t=0$), the population
imbalances $m$ exhibit step behaviors for the finite-size lattice strengths.
The corresponding step gap is given by $\Delta =2/(Ln)$. When the lattice
strength increases, this step gap becomes small and especially $\Delta
\rightarrow 0$ for $L\rightarrow \infty $. In the presence of SOC ($\lambda
/t\neq 0$), the finite-size step behaviors still exist but become smoother, since SOC
can make fermions hop between the nearest-neighbor sites with spin flipping
and thus has a strong effect on the population imbalances $m$. Similarly,
the finite-size step behaviors also exist when the population imbalances $m
$ vary as the SOC strength; see the insets of Figs.~\ref{fig9}(b) and~\ref%
{fig9}(d). In terms of Eq.~(\ref{eq-Q}), we find straightforwardly that the
finite-size step behaviors with respect to the SOC strength can lead to the
similar behaviors of the center-of-mass momentum $Q$; see Figs.~\ref{fig9}%
(b) and~\ref{fig9}(d). Apart from the finite-size effect, the step behaviors of the
center-of-mass momentum $Q$ depend strongly on the Zeeman field and the SOC
strength. For some parameter regimes, we find numerically that the corresponding steps become unobvious; see, for example, the red dash-dot line for $h/t=1.5$ and $L=60$ in Figs.~\ref{fig9}(b).

\begin{figure}[t]
\centering\includegraphics[width = 6.0in]{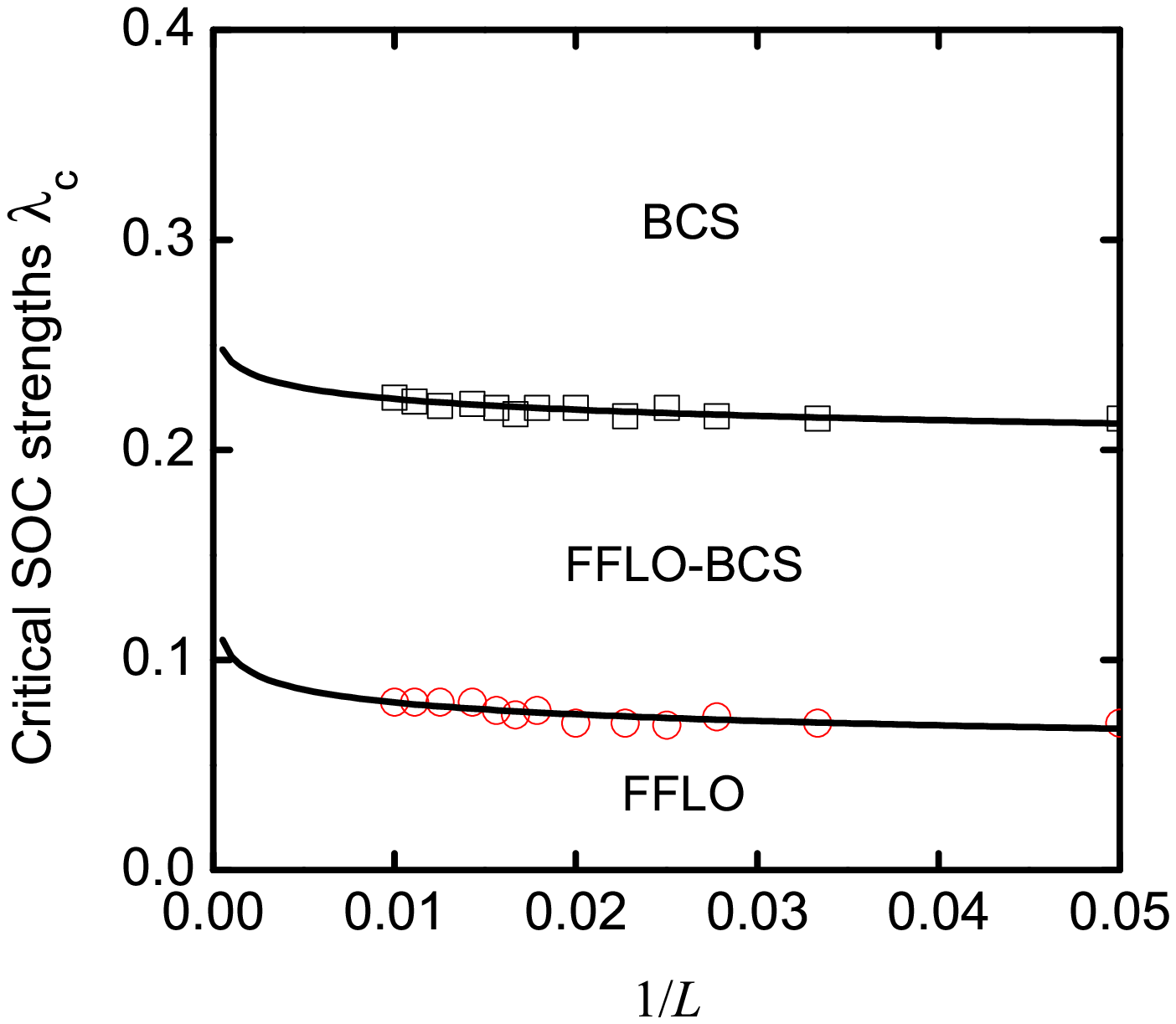}
\caption{\textbf{The critical SOC strengths }$\protect\lambda _{c}$\textbf{\
as functions of the lattice length.}}
\label{fig10}
\end{figure}

It should be pointed out that the boundary condition may influence the spin
polarizations at the two ends; however, it does not affect our main
predictions about spin polarizations in both real and momentum spaces, as
demonstrated in Figs.~\ref{fig4} and \ref{fig5} with $L=60$ and $L=100$. We
also do not observe phase separation in the open boundary condition. So we
can exclude the possibility of three peaks in the FFLO-BCS phase from the
phase-separation effect. In Fig.~\ref{fig10}, we plot the critical SOC
strengths $\lambda _{c}$, which govern the phase boundaries, as functions of
the lattice length, when the Zeeman field $h/t=1.5$ (the dash line of Fig.~%
\ref{fig12}). In terms of this finite-size-scaling analysis, we find that
when increasing the lattice length, our predicted FFLO-BCS phase, with a
unique three-peak structure, still exists, although the center-of-mass
momentum $Q$ and the phase boundaries change slightly.

\subsection{Phase diagram.}

Having identified three superfluid phases, including the FFLO-BCS, FFLO, and
BCS phases, we now figure out the corresponding phase diagram as a function
of the filling factor, the Zeeman field, and the SOC strength. Numerically,
the FFLO-BCS, FFLO, and BCS phases are characterized by three, two, and one
peak(s) in the pairing momentum distribution $P(k)$, respectively. In
addition, when the fermions are fully polarized, i.e., $m=1$, no pairing can
occur. The corresponding phase is referred as the fully-polarized (FP) phase%
\cite{YAL10}. The boundary between the FFLO and FFLO-BCS phases can be
determined by Eq.~(\ref{B1}), whereas the boundary between the FFLO-BCS and
BCS phases can be determined by Eq.~(\ref{B2}).

\begin{figure}[t]
\centering\includegraphics[width = 4.5in]{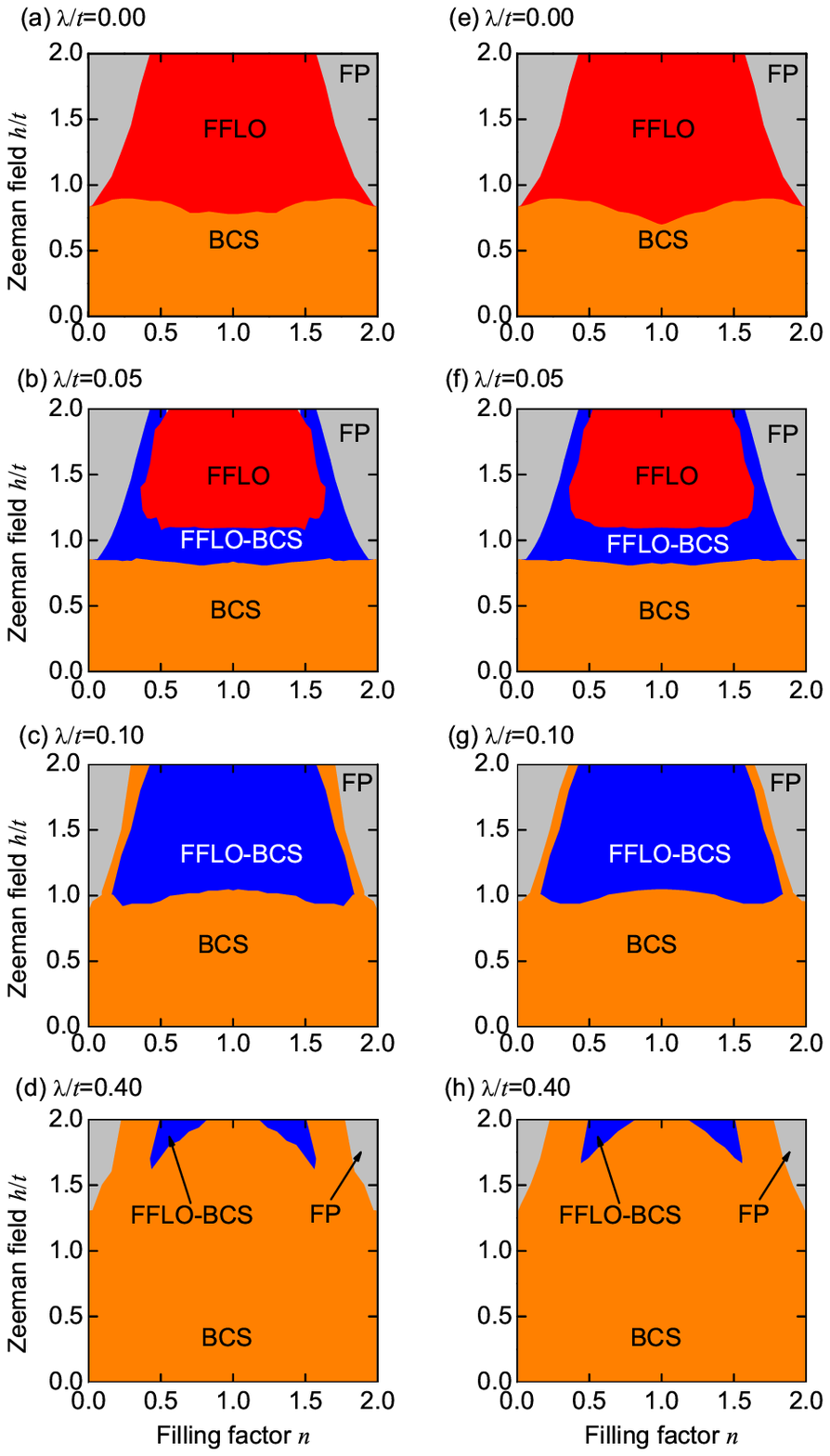}
\caption{\textbf{Phase diagrams in the }$h-n$\textbf{\ plane for the
different SOC strengths.}}
\label{fig11}
\end{figure}

\begin{figure}[t]
\centering\includegraphics[width = 6.0in]{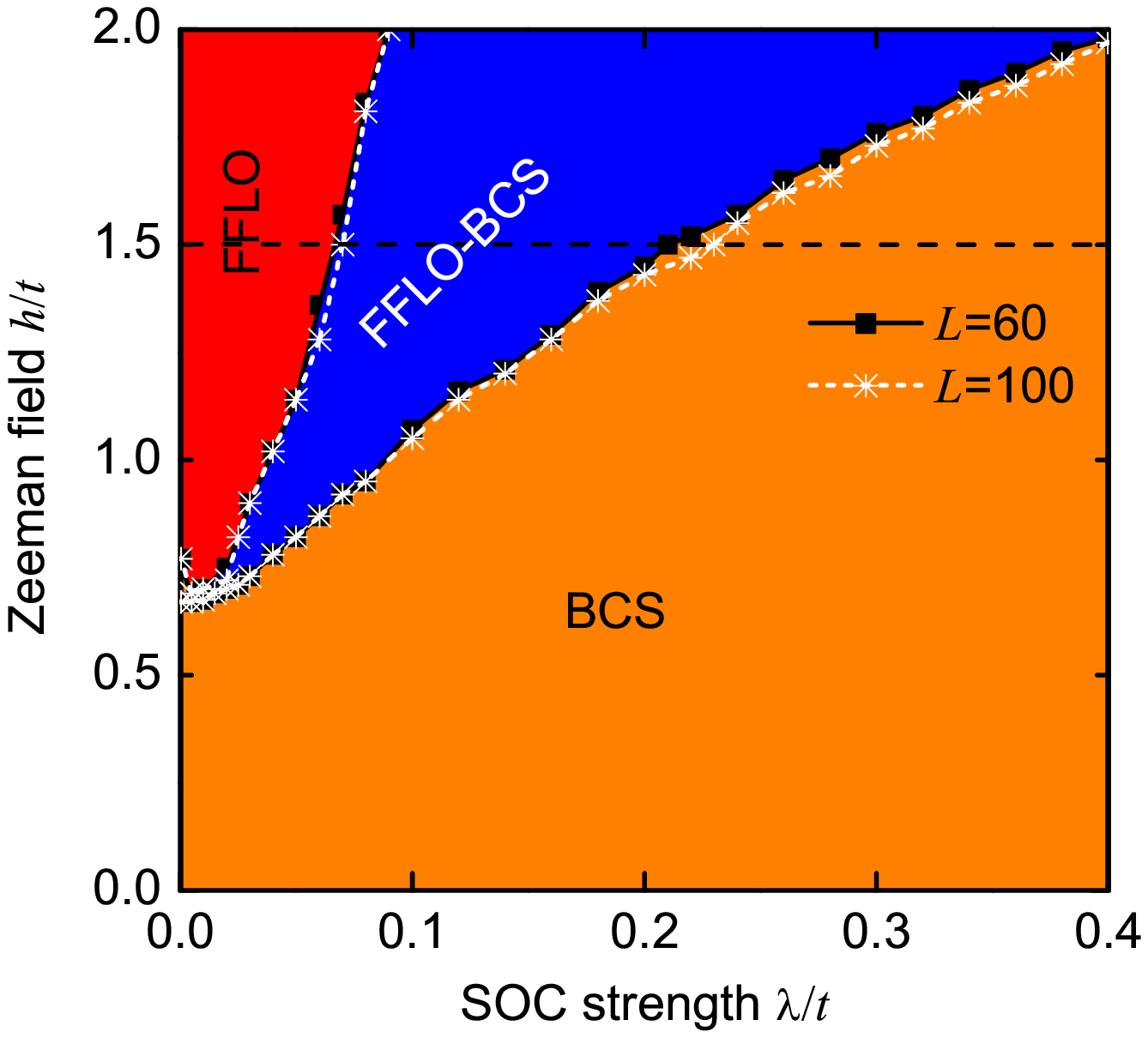}
\caption{\textbf{Phase diagram in the }$h-\protect\lambda $\textbf{\ plane
for the different lattice lengths }$L=60$\textbf{\ and }$L=100$\textbf{. }}
\label{fig12}
\end{figure}

In Fig.~\ref{fig11}, we plot the phase diagram in the $n-h$ plane. In the
absence of SOC ($\lambda /t=0$), the FP, BCS, and FFLO phases can be found%
\cite{ML07}; see Fig.~\ref{fig11}(a). For the weak SOC strength (see, for
example, $\lambda /t=0.05$), the FFLO-BCS phase can be found, and the FFLO
phase is suppressed; see Fig.~\ref{fig11}(b). When the SOC strength $\lambda
/t=0.1$, the FFLO phase vanishes, and the FFLO-BCS phase is enhanced; see
Fig.~\ref{fig11}(c). For the strong SOC strength (see, for example, $\lambda
/t=0.4$), the FFLO-BCS phase almost disappears, and the BCS phase dominates
in the whole parameter regime; see Fig.~\ref{fig11}(d). These results
demonstrate that for the weak SOC strength, a large regime for the FFLO phase can
always be observed. However, for the strong SOC strength, the interband FFLO
pairing are completely suppressed and the intraband BCS pairing always
dominates in the whole parameter regime. This result is in contrast to that
from mean-field prediction\cite{QU1D}, in which the FFLO phase always exists
even for a stronger SOC strength ($\lambda /t>1$).

In addition, all the phase diagrams in Fig.~\ref{fig11} are symmetric about
the half filling ($n=1$), which can be understood from the following
particle-hole transformation:
\begin{equation}
c_{i\uparrow }\rightarrow -(-1)^{i}d_{i\downarrow }^{\dagger },\quad
c_{i\downarrow }\rightarrow (-1)^{i}d_{i\uparrow }^{\dagger }.  \label{TS}
\end{equation}%
Under the transformation (\ref{TS}), we find (see Methods section)
\begin{equation}
\mathcal{H}(t,\mu ,h,U,\lambda )\rightarrow \mathcal{H}(t,-\mu ,h,U,\lambda
),  \label{PHS}
\end{equation}%
Here we have introduced a chemical potential $\mu $ to the original
Hamiltonian (\ref{H1}), which equals exactly to zero at the half filling.
Equation (\ref{PHS}) demonstrates that the Hamiltonian (\ref{H1}) has the
particle-hole symmetry. This symmetry ensures that the relevant physics in
the low filling factor regime ($n<1$) is identical to that in the high
filling factor regime ($n>1$), i.e., we have the observation in Fig.~\ref%
{fig11}.

Figure~\ref{fig12} shows the phase diagram in the $h-\lambda $ plane at the
half filling ($n=1$), which further confirms that the interband FFLO pairing
can be suppressed by the intraband BCS pairing. However, the situation for
the FFLO-BCS phase is quite different. Since this phase requires not only an
appropriate spin polarization but also a finite energy difference between $%
\varepsilon _{\text{F}\pm }$ (see Supplementary Information), we see that it
is more likely to be observed at a finite SOC strength and a stronger Zeeman
field. Obviously, without the Zeeman field ($h/t=0$), the spin is fully
polarized along $x$ direction ($\theta _{\pm }=\pi /2$), and only BCS phase
can be observed. In the presence of SOC, a stronger Zeeman field is thus
required to bring the polarization along $z$ direction (smaller than the
critical polarization angle), so as to favor the interband FFLO pairing. We
choose the results of $h/t=1.5$ as an example to illustrate this point. In
the absence of SOC ($\lambda /t=0$), the polarization angles $\theta _{\pm
}=0$, and we can only observe the FFLO phase. When $\lambda /t<0.07$ ($%
\theta _{\pm }\simeq \pi /40$), this interband FFLO pairing always
dominates. However, when $0.07<\lambda /t<0.21$ ($\theta _{\pm }\simeq \pi
/13$), we find the FFLO-BCS phase. Finally, when $\lambda /t>$ $0.21$, the
intraband BCS pairing dominates. Strikingly, we find that these critical
angles are generally of the order of $\pi /10$, thus it is very easily to
drive the FFLO phase to the BCS phase by a weak SOC. \newline

\section*{{\protect\LARGE \textbf{Discussion}}}

\begin{figure}[t]
\centering\includegraphics[width = 5.5in]{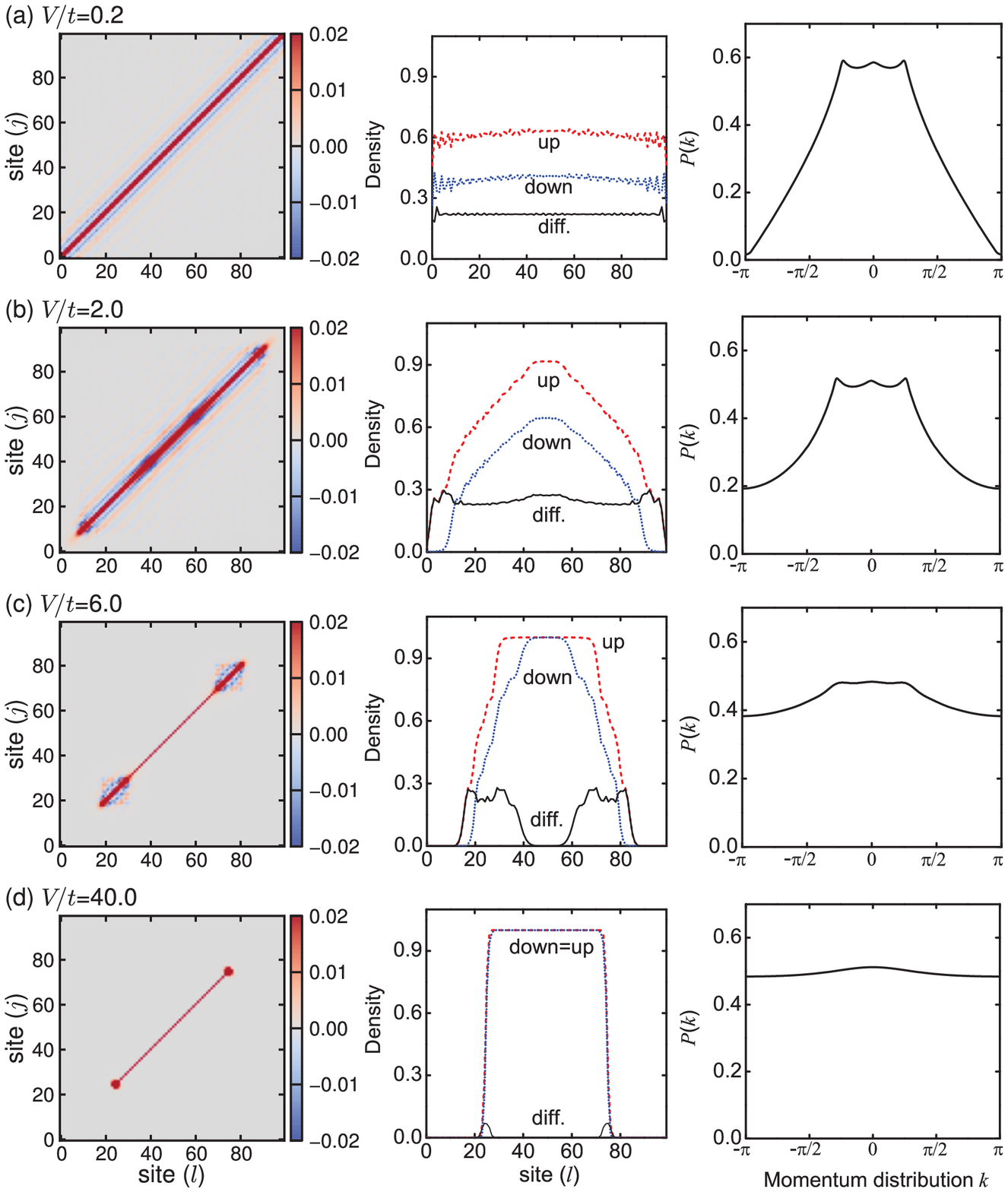}
\caption{\textbf{The pairing correlation functions }$P(l,j)$\textbf{\ (left
column), the local densities }$n(l)$\textbf{\ (center column), and the
pairing momentum distributions }$P(k)$\textbf{\ (right column) for the
different trapped frequencies.}}
\label{fig13}
\end{figure}

In real experiments, a harmonic trapped potential usually exists, and the
Hamiltonian (\ref{H1}) should be added an extra term
\begin{equation}
\mathcal{H}_{\text{trap}}=V\left( \frac{2}{L-1}\right)
^{2}\sum\limits_{l}\left( l-\frac{L+1}{2}\right) ^{2}n_{l},  \label{TRAP}
\end{equation}%
where $V$ is the trapped frequency. In Fig.~\ref{fig13}, we plot the pairing
correlation functions $P(l,j)$, the local densities $n(l)$, and the pairing
momentum distributions $P(k)$ for the different trapped frequencies, when $%
h/t=1.5$, $\lambda /t=0.16$, and $L=100$. The results for the other lattice
length (such as $L=60$) are similar and thus not plotted here.

Without the trapped potential, the system is located at the FFLO-BCS phase,
in which the pairing correlation function $P(l,j)$ has an oscillating
behavior and the pairing momentum distribution $P(k)$ has a unique
three-peak structure; see Figs.~\ref{fig5} and~\ref{fig7}. For a weaker
trapped frequency (see, for example, $V/t=0.2$), the oscillation of the
pairing correlation function $P(l,j)$ and especially three peaks of the
pairing momentum distribution $P(k)$ still exist; see Fig.~\ref{fig13}(a).
In addition, the corresponding density profile is almost the same as that
without trapped potential; see Fig.~\ref{fig3}(f). It means that no
obvious phase separation in real space occurs. Thus, the predicted phase
diagrams in Figs.~\ref{fig11} and~\ref{fig12}, including the FFLO-BCS phase,
also remain, although the corresponding phase boundaries change slightly.
However, due to the existence of the trapped potential, the particle-hole
symmetry of the inhomogeneous Hamiltonian $\mathcal{H}+\mathcal{H}_{\text{%
trap}}$ is broken, and the phase diagrams in Fig.~\ref{fig11} are not
symmetric about the half filling ($n=1$). When increasing the trapped
frequency (see, for example, $V/t=2.0$ and $6.0$), the phase separation in
real space occurs, since in this case the number of the fermions in the
different sites is not same\cite%
{AF07,MT08,MT10,Martin2006,Guthrie2006,MR03,HM10}. When the trapped
frequency $V/t=2.0$, the pairing correlation function $P(l,j)$ exhibits an
oscillation in $5<l<55$, and the pairing momentum distribution $P(k)$ has
three peaks. However, the local density $n(l)$ shows that the sites are
fully polarized in two sides. It means that the FFLO-BCS and FP phases are
mixed, and the system is thus located at the FFLO-BCS phase core with the FP
phase wings; see Fig.~\ref{fig13}(b). When the trapped frequency $V/t=6.0$,
the oscillation regime of the pairing correlation function $P(l,j)$ turns
into $10<l<20$ and $40<l<50$, and moreover, the pairing momentum
distribution $P(k)$ becomes smoother, i.e., no obvious peaks can be found.
In addition, the local density $n(l)$ shows the emergence of five phases,
including the vacuum, FP, partly-polarized, metal, and band insulator phases
(from left to center of the lattice); see Fig.~\ref{fig13}(c). In the metal
phase, all spin-down fermions can move freely in a uniform background of the
spin-up fermions, and the band insulator is fully occupied by the spin-up
and spin-down fermions\cite{FG97}. Due to the phase separation in real
space, the phase boundaries and the phase diagrams are hardly to be
determined\cite{AF07}. For a large trapped frequency (see, for example, $%
V/t=40.0$), the physics is quite different, since in such case the term $%
\mathcal{H}_{\text{trap}}$ dominates in the inhomogeneous Hamiltonian. As a
consequence, all fermions are forced to the center of the trap and there is
only the band insulator without any moving fermions; see Fig.~\ref{fig13}%
(d). From above discussions, it can be seen that our predictions could be
observed for a weaker trapped potential, which is easily prepared in
experiments.

In summary, we have shown, using the state-of-the-art DMRG calculations,
that the true pairings in a 1D optical lattice can be completely modified by
SOC, due to the induced triplet pairing. Especially, this system admits an
exotic coexistence of the interband FFLO and intraband BCS pairings for the
weak and moderate SOC strengths. However, for the strong SOC strength, the
intraband BCS pairing always dominates, and the relevant physics is thus the
BCS superfluid in the whole parameter regime. This yields a new picture to
understand the true pairings in 1D spin-orbit coupled degenerate Fermi
gases. The last conclusion (III) should be useful for searching the
topological superfluids in this model. Finally, we have addressed the effect
of the trapped potential on the pairing correlations and the local density.
We have shown that our predictions could be observed in a weaker trapped
potential, which is easily prepared in experiments.

\section*{{\protect\LARGE \textbf{Methods}}}

By means of the transformation (\ref{TS}), we find that the kinetic energy $%
\sum_{is}c_{is}^{\dagger }c_{is}\rightarrow
-\sum_{is}d_{is}d_{i+1s}^{\dagger }=\sum_{i}d_{is}^{\dagger }d_{is}$, the
chemical potential and Zeeman field $\mu (n_{i\uparrow }+n_{i\downarrow
})+h(n_{i\uparrow }-n_{i\downarrow })\rightarrow 2\mu -\mu (d_{i\downarrow
}^{\dagger }d_{i\downarrow }+d_{i\uparrow }^{\dagger }d_{i\uparrow
})-h(d_{i\uparrow }^{\dagger }d_{i\uparrow }-d_{i\downarrow }^{\dagger
}d_{i\downarrow })$, the on-site attractive interaction $(n_{i\uparrow }-{%
\frac{1}{2}})(n_{i\downarrow }-{\frac{1}{2}})\rightarrow (1-d_{i\uparrow
}^{\dagger }d_{i\uparrow }-{\frac{1}{2}})(1-d_{i\downarrow }^{\dagger
}d_{i\downarrow }-{\frac{1}{2}})=(d_{i\uparrow }^{\dagger }d_{i\uparrow }-{%
\frac{1}{2}})(d_{i\downarrow }^{\dagger }d_{i\downarrow }-{\frac{1}{2}})$,
and the SOC term $c_{l\uparrow }^{\dagger }c_{l+1\downarrow }-c_{l\downarrow
}^{\dagger }c_{l+1\uparrow }+c_{l+1\downarrow }^{\dagger }c_{l\uparrow
}-c_{l+1\uparrow }^{\dagger }c_{l\downarrow }\rightarrow
-(-1)^{2l+1}(d_{l\downarrow }d_{l+1\uparrow }^{\dagger }-d_{l\uparrow
}d_{l+1\downarrow }^{\dagger }+d_{l+1\uparrow }d_{l\downarrow }^{\dagger
}-d_{l+1\downarrow }d_{l\uparrow }^{\dagger })=d_{l\uparrow }^{\dagger
}d_{l+1\downarrow }-d_{l\downarrow }^{\dagger }d_{l+1\uparrow
}+d_{l+1\downarrow }^{\dagger }d_{l\uparrow }-d_{l+1\uparrow }^{\dagger
}d_{l\downarrow }$. As a result, we derive $\mathcal{H}(t,\mu ,h,U,\lambda
)\rightarrow \mathcal{H}(t,-\mu ,h,U,\lambda )$, i.e., Eq.~(\ref{PHS}).%
\newline

\section*{{\protect\LARGE \textbf{Acknowledgments}}}

We acknowledge Profs.~Wei Yi, An-chun Ji, and Qing Sun, and Dr.~Chunlei Qu
for their valuable discussions. This work is supported partly by the 973
program under Grant No.~2012CB921603; the NNSFC under Grants No.~11422433
and No.~61275211; the NCET under Grant No.~13-0882; the FANEDD under Grant
No.~201316; the OIT under Grant No.~2013804; OYTPSP; and SSCC. M.G. is
supported by Hong Kong RGC/GRF Projects (No.~401011 and No.~2130352),
University Research Grant (No.~4053072) and The Chinese University of Hong
Kong (CUHK) Focused Investments Scheme.

\section*{{\protect\LARGE \textbf{Author Contributions}}}

S.G., M.G., G.C., and S.J. conceived the idea, J.L. and X.Z. performed the
numerical calculations, C.P.H. and K.Z. performed the theoretical
calculations, M.G. and G.C. wrote the manuscript, M.G., G.C. and S.J.
supervised the whole research project. J.L. and X.Z. contributed equally to
this work.

\section*{{\protect\LARGE \textbf{Additional information}}}

\textbf{Supplementary information} accompanies this paper at
http://www.nature.com/scientificreports \newline
\textbf{Competing financial interests:} The authors declare no competing
financial interests.

\newpage

\textbf{Figure 1: (a) The scaled ground-state energy }$E_{g}/(Lt)$\textbf{\
and (b) the truncation error as functions of the number of states kept (SK).}
In all subfigures, $n=1$, $\lambda /t=0.16$, and $h/t=1.5$.

\textbf{Figure 2: A schematic picture for illustrating unconventional
pairings in a 1D optical lattice.} (a) Without SOC, (b) the weak and
moderate SOC strengths, and (c) the strong SOC strength. When both bands are
partially occupied, there are two Fermi surfaces, denoted by $\varepsilon _{%
\text{F}\pm }$, which give rise to four Fermi points $\pm k_{1}$ and $\pm
k_{2}$ (see Supplementary Information). The corresponding spin-polarized
angles at the two Fermi surfaces are defined as $\theta _{\pm }$, which are
determined by the SOC strength and the Zeeman field; see Eq.~(\ref{SPA}).
These polarizations are essential for describing the true pairings of the
Hamiltonian (1). In (a), the spin is fully polarized along $z$ direction ($%
\theta _{\pm }=0$), and only the interband FFLO pairing is thus formed. For
the weak and moderate SOC strengths, $\theta _{\pm }$ are typically of the
order of $\pi /10$ (see Supplementary Information). In such case, both the
interband FFLO and intraband BCS pairings are allowed; see (b). More
importantly, these two pairings can coexist, leading to a new phase called
the FFLO-BCS phase. This new phase is characterized by a unique three-peak
structure in pairing momentum distribution. For the strong SOC strength, the
spin is almost polarized along $x$ direction ($\theta _{\pm }\sim \pi /2$),
and the intraband BCS pairing thus dominates; see (c).

\textbf{Figure 3: The pairing correlation functions }$P(l,j)$\textbf{\ and
local densities }$n(l)$\textbf{\ for the different SOC strengths. }Left two
columns for $L=60$ and right two columns for $L=100$. In the local density,
the solid line marks the local spin difference (diff.), which is defined as $%
s_{z}=\langle n_{l\uparrow }-n_{l\downarrow }\rangle $. In all subfigures, $%
n=1$ and $h/t=1.5$.

\textbf{Figure 4: The off-diagonal pairing correlation functions }$P(l,L-l)$%
\textbf{.} (a) $P(l,L-l)$\ for the different SOC strengths, when $L=60$, $n=1
$, and $h/t=1.5$. (b) shows the zoomed images of the center $20$\ sites of
(a). (c) and (d) are the same as those of (a) and (b), but with $L=100$.

\textbf{Figure 5: The pairing momentum distributions }$P(k)$\textbf{. }(a) $%
P(k)$ for the different SOC strengths, when $L=60$, $n=1$, and $h/t=1.5$.
(b) shows the zoomed image of (a). (c) and (d) are the same as those of (a)
and (b), but with $L=100$.

\textbf{Figure 6: The ground-state energy }$E_{g}/t$\textbf{\ as a function
of the SOC strength. }(a) $L=60$ and (b) $L=100$. In all subfigures, $n=1$
and $h/t=1.5$.

\textbf{Figure 7: The center-of-mass momentum }$Q$\textbf{.} (a) $Q$, which
is derived respectively from the state-of-the-art DMRG calculations
(Symbols) and analytical Eq.~(\ref{eq-Q}) (Solid line), as a function of the
SOC strength, when $h/t=1.5$, $n=1$, and $L=60$. In the analytical result,
the population imbalance $m$ is also obtained from the state-of-the-art DMRG
calculations. (b) $Q$ as a function of the population imbalance, when $%
\lambda /t=0.06$, $n=1$, and $L=60$. (c) and (d) are the same as those of
(a) and (b), but with $L=100$.

\textbf{Figure 8: The critical population imbalance }$m_{c}$\textbf{\ as a
function of the SOC strengths}. In this figure, $n=1$.

\textbf{Figure 9: The population imbalance }$m$ \textbf{and the
center-of-mass momentums }$Q$ \textbf{for the different Zeeman fields.} (a) $%
m$ as a function of the Zeeman field for the different SOC strengths, when $%
n=1$ and $L=60$. (b) $Q$ as a function of the SOC strength for the different
Zeeman fields, when $n=1$ and $L=60$. The inset of (b) shows $m$ as a
function of the SOC strength. (c) and (d) are the same as those of (a) and
(b), but with $L=100$.

\textbf{Figure 10: The critical SOC strengths }$\lambda _{c}$\textbf{\ as
functions of the lattice length. }In this figure, $n=1$ and $h/t=1.5$.

\textbf{Figure 11: Phase diagrams in the }$h-n$\textbf{\ plane for the
different SOC strengths. }(a)-(d) $L=60$\ and (e)-(h) $L=100$\textbf{.}

\textbf{Figure 12: Phase diagram in the }$h-\lambda $\textbf{\ plane for the
different lattice lengths }$L=60$\textbf{\ and }$L=100$\textbf{. }In this
figure, $n=1$.

\textbf{Figure 13: The pairing correlation functions }$P(l,j)$\textbf{\
(left column), the local densities }$n(l)$\textbf{\ (center column), and the
pairing momentum distributions }$P(k)$\textbf{\ (right column) for the
different trapped frequencies.} In this figure, $h/t=1.5$, $\lambda /t=0.16$%
, and $L=100$.

\end{document}